\documentclass[useAMS,usenatbib,usegraphicx]{mn2e}
\usepackage[dvipsnames]{color} 

\newcommand{\be}{\begin{equation}}
\newcommand{\ee}{\end{equation}}

\newcommand{\msun}{\mbox{M$_{\sun}$}}

\newcommand{\bl}[1]{\mbox{\boldmath$ #1 $}}
\newcommand{\rhat}{\mbox{\boldmath$ \hat{r}$}} 
\newcommand{\phihat}{\mbox{\boldmath$ \hat{\phi}$}} 

\newcommand{\mdot}{\dot{M}}

\newcommand{\Qcr}{Q_{\rm cr}}
\newcommand{\ul}{\underline{\hspace{20pt}}}
\def\la{\mathrel{\hbox{\rlap{\hbox{\lower4pt\hbox{$\sim$}}}\hbox{$<$}}}}
\def\ga{\mathrel{\hbox{\rlap{\hbox{\lower4pt\hbox{$\sim$}}}\hbox{$>$}}}}
\begin{document}

\title[Protostellar disc accretion]{
Self-regulated gravitational accretion in protostellar discs}
\author[E. I. Vorobyov and S. Basu]{E. I. Vorobyov$^{1,2}$\thanks{E-mail:
vorobyov@astro.uwo.ca (EIV); basu@astro.uwo.ca (SB)} and Shantanu
Basu$^{1}$ \\
$^{1}$Department of Physics and Astronomy, University of Western Ontario, 
London, Ontario, N6A 3K7, Canada \\
$^{2}$ Institute of Physics, South Federal University, Stachki 194, Rostov-on-Don, Russia.}

\date{}

\maketitle

\label{firstpage}

\begin{abstract}
We present a numerical model for the evolution of a protostellar
disc that has formed self-consistently from the collapse of a molecular
cloud core. The global evolution of the disc is followed for several 
million years 
after its formation. The capture of a wide range of spatial
and temporal scales is made possible by use of the thin-disc approximation. 
We focus on the role of gravitational torques in transporting 
mass inward and angular momentum outward during different evolutionary
phases of a protostellar disc with disc-to-star mass ratio of 
order 0.1. In the early phase, when the
infall of matter from the surrounding envelope is substantial, mass is
transported inward by the gravitational torques from spiral arms that are
a manifestation of the envelope-induced gravitational instability in
the disc. In the late phase, when the
gas reservoir of the envelope is depleted, the distinct spiral structure 
is replaced by ongoing irregular nonaxisymmetric density perturbations. The 
amplitude of these density perturbations decreases with time, 
though this process is moderated 
by swing amplification 
aided by the existence of the disc's sharp
outer edge.
Our global modelling of the protostellar disc reveals that
there is typically a residual nonzero gravitational torque from these density
perturbations, i.e. their effects do not exactly cancel out in each region.
In particular, the net gravitational torque in the inner disc tends
to be negative during first several million years of the evolution, while
the outer disc has a net positive gravitational
torque. 
Our global model of a self-consistently formed disc shows that it is
also self-regulated in the late phase, so that it is near the 
Toomre stability limit, with a near-uniform Toomre parameter 
$Q\approx 1.5-2.0$. Since the disc also has near-Keplerian rotation, 
and comparatively weak temperature variation, it
maintains a near-power-law surface density profile proportional to
$r^{-3/2}$.
\end{abstract}


\section{Introduction}

The presence of finite angular momentum in a prestellar cloud core
provides a significant obstacle to the formation of stellar-sized objects, contributing
to the overall ``angular momentum problem'' of star formation. Even the
relatively small ($\sim 1$ km s$^{-1}$ pc$^{-1}\, \approx 10^{-14}$ rad s$^{-1}$)  
rotation rates measured in cloud cores \citep[e.g.][]{Goodman} imply that
most of the infalling matter will land on a protostellar disc rather than directly on to
the protostar. Indeed, discs are observed
or inferred around at least some class 0 and class I protostars, most T Tauri
stars, and even around brown dwarfs. However, the observed mass ratios
between the disc and central object are typically $\sim 1\%$ 
\citep{Andrews, Scholz1}. This implies
that there is an efficient mechanism to transport angular momentum outward
and mass inward that sets in very early during the life of the disc and even
while it is forming.

\citet{VB1,VB2} have recently modeled numerically the 
self-consistent collapse of a rotating molecular cloud leading to
the formation and evolution of a protostellar disc. During the early
($< 0.5$ Myr) evolution, 
mass infall from the core envelope can episodically destabilise
the disc through gravitational instability and lead to the formation 
of spiral structure and dense clumps within the arms. 
Gravitational torques associated with the spiral arms drive the
clumps on to the protostar, generating mass accretion and luminosity bursts
comparable to those observed in FU Ori stars.
During this early phase, most of the protostellar mass is accreted during
the bursts rather than in the quiescent phase between the bursts.
The early burst phase terminates when the infalling envelope has lost
most of its gas reservoir. The protostar then enters a T Tauri phase and 
its subsequent evolution is characterised by a low-level accretion.
The physical mechanism or mechanisms that drive this low-level accretion
are not well understood. 
The magnetorotational instability \citep[MRI;][]{Balbus} 
has been suggested as the means of angular momentum and mass transport 
in protostellar discs. However,   
the thermal ionisation in most parts of T Tauri discs is
too low to allow sufficient magnetic coupling for the MRI to operate.
This problem may
be mitigated by invoking nonthermal ionisation by cosmic rays
in the upper layers of the disc \citep{Gammie96}, or X-rays from the central 
star \citep{Glassgold}. \citet{Turner} have shown that the MRI-induced 
activity can indeed extend to the disc midplane at distances $\sim 5$ AU 
under some circumstances. Nevertheless, the overall ability 
of partial unstable
regions of the disc to drive effective transport throughout the disc
remains unresolved. Furthermore, \citet{Hartmann} have recently
suggested that a combination of the MRI and
gravitational instabilities may be necessary to explain the observed 
accretion rates for at least the more massive T Tauri stars.

Given the lingering uncertainties associated with the MRI and 
the additional suggestion made by \citet{Hartmann} that disc masses may have been
systematically underestimated (due to poorly constrained dust opacities),
it is tempting to consider the T Tauri accretion phase 
as merely the residual phase of gravitationally driven accretion.
Numerical simulations of {\it isolated} protostellar discs do indicate that 
gravitational instabilities may work if the disc masses are sufficiently large 
\citep[e.g.][]{Tomley, Laughlin, Lodato05}, and
efforts have also been made to describe the effect of such global 
instabilities in terms of local viscous 
dynamics \citep[e.g.][]{Lodato04}.
Numerical studies with simple prescribed cooling and heating have shown that the
nonaxisymmetric structure in protostellar discs washes out and density 
fluctuations become less than one per cent when the Toomre $Q$-parameter 
becomes larger than a few \citep[e.g.][]{Pickett, Lodato07}. 
However, it is not clear when this phase begins and gravitational 
instabilities cease to operate.

In our opinion, the fundamental limitation of the above numerical
simulations is in the isolated nature of the model discs.
The onset of the axisymmetric phase (if any) should depend not only on the disc physics 
but also on the physics of the surrounding medium. The infall of gas from the
surrounding envelope (certainly in the early evolutionary phase) or 
gravitational perturbations from companions and parent nonaxisymmetric molecular 
cloud can drive a protostellar disc away from an axisymmetric state.  
A memory of initial core conditions also determines the distribution
of angular momentum and mass in the disc as well as its intrinsic size. 
Edge-effects can certainly affect the disc's ability to support 
persistent density fluctuations, as we emphasise in this paper.
As \citet{Larson} pointed out, even small density fluctuations of the order 
of a few per cent in a self-gravitating disc 
can create gravitational torques that can provide angular momentum transport
comparable to what is often invoked via the 
ad-hoc $\alpha$-viscosity mechanism. 
Observations do seem to support the existence of a pronounced nonaxisymmetry
in discs that are several Myr old, e.g. around AB Aurigae \citep{Fukagawa} and 
HD~100546 \citep{Grady}.
Hence, it is important for numerical models to take into account the effects of the environment
on the long-term evolution of protostellar discs.

In this paper we extend our previous numerical analysis of disc accretion
\citep{VB1,VB2} {\it to the T Tauri phase}. We start our
simulation in the prestellar phase and terminate it when the central protostar 
and the surrounding disc are about 3~Myr old. Such long integration times
are made possible by the use of the thin-disc approximation. 
We are particularly interested in the ability of gravitational torques to drive accretion 
in relatively aged protostellar discs. Here, we focus on the evolution 
of a single model in order to present a detailed study of the internal 
disc dynamics. A parameter study is left for future presentation.

\section{Model description and initial conditions}
\label{model}

We use the thin-disc approximation to compute the evolution of nonaxisymmetric rotating, 
gravitationally bound cloud cores. For details of the basic equations, numerical methods,
and numerical tests we refer the reader to \citet{VB2}. 
Here we briefly provide the basic equations. For simplicity, we neglect the
contribution of a frozen-in supercritical magnetic field 
\citep[accounted for in some models of][]{VB2}, which does not change the
main qualitative results.
The equations of mass and momentum transport are

\begin{eqnarray}
\label{cont}
 \frac{{\partial \Sigma }}{{\partial t}} & = & - \nabla _p  \cdot \left( \Sigma \bl{v}_p \right), \\ 
\label{mom}
 \Sigma \frac{d \bl{v}_p }{d t}  & = &  - \nabla _p {\cal P}  + \Sigma \bl{g}_p \, ,
\end{eqnarray}
where $\Sigma$ is the mass surface density, 
${\cal P}$ is the vertically integrated gas pressure,
$\bl{v}_p = v_r \rhat + v_{\phi} \phihat$ is the velocity in the
disc plane, $\bl{g}_p = g_r \rhat + g_{\phi}\phihat$ is the gravitational acceleration in the disc plane,
and $\nabla_p = 
\rhat \partial/\partial r + \phihat \, r^{-1}  \partial/\partial \phi$ 
is the gradient along the planar coordinates of the disc.
The system of equations is closed with a barotropic equation
that makes a transition from isothermal to adiabatic evolution at a
critical density, i.e. 
\begin{equation}
{\cal P}=c_{\rm s}^2\Sigma
+ c_{\rm s}^2 \Sigma_{\rm cr} \left(\Sigma \over \Sigma_{\rm cr} \right)^\gamma,
\label{eos}
\end{equation}
where $c_{\rm s}$ is the isothermal sound speed, $\gamma=7/5$ is the ratio of specific
heats, and $\Sigma_{\rm cr}=36.2$~g~cm$^{-2}$, which corresponds to a critical 
number density $n_{\rm cr} = 10^{11}$ cm$^{-3}$ under the assumption of vertical
hydrostatic equilibrium \citep[see][]{VB2}. Equation~(\ref{eos}) yields
a density-temperature relation that is in good agreement with the density-temperature
relation derived using exact spherically symmetric frequency-dependent
radiation transfer simulations \citep[for a detailed comparison see][]{VB2}.
We have also not included the effect of stellar irradiation, which can be a
significant heating source in the disc during the late accretion phase \citep{Chiang}. 
Such considerations require detailed modeling of (or assumptions about) 
the vertical structure (e.g. flaring) of the disc \citep[see also][]{Garaud}. 
For simplicity, we do not
keep track of the vertical structure of the disc in this study. 
The gravitational potential in the plane of the disc is computed as
\begin{eqnarray}
 \Phi(r,\phi,z=0,t) &=&  - G \int_0^{r_{\rm out}} r^\prime dr^\prime   
  \nonumber \\
      & &      \times  \int_0^{2\pi} 
               \frac{\Sigma(r^\prime,\phi^\prime) d\phi^\prime}
                    {\sqrt{{r^\prime}^2 + r^2 - 2 r r^\prime
                       \cos(\phi^\prime - \phi) }}  \, ,
\label{potential}
\end{eqnarray}
where $r_{\rm out}$ is the radius of the cloud core 
\citep[see][]{BT}.

Equations~(\ref{cont})--(\ref{potential}) are solved in polar coordinates
$(r, \phi)$ on a numerical grid with
$128 \times 128$ points. The radial points are logarithmically spaced.
The innermost grid point is located at $r=5$~AU, and the size of the 
first adjacent cell is 0.3~AU. We initiate the accretion phase after 
the central surface density exceeds $\Sigma_{\rm cr}$ by introducing a ``sink cell'' at $r<5$~AU, 
which represents the central protostar plus some circumstellar disc material, 
and impose a free inflow inner boundary condition. 
The outer boundary (at 8000~AU) remains fixed in position and there is no radial inflow or outflow
allowed, i.e. the cloud has a constant mass and volume. 

In this paper we analyse the details of the internal disc dynamics for a single
model which follows the evolution of a cloud core with initial mass
$M_{\rm cl}=0.8~M_\odot$ and outer radius $r_{\rm out}=0.04$~pc.
The gas has a mean molecular mass $2.33 \, m_{\rm H}$ and is initially
isothermal with temperature $T=10$~K. The subsequent high-density evolution is of course
not isothermal due to our barotropic equation of state. 
The initial surface density ($\Sigma$) and angular velocity ($\Omega$) distributions 
are characteristic of a collapsing axisymmetric magnetically
supercritical core \citep{Basu}:
\begin{equation}
\Sigma={r_0 \Sigma_0 \over \sqrt{r^2+r_0^2}}\:,
\label{dens}
\end{equation}
\begin{equation}
\Omega=2\Omega_0 \left( {r_0\over r}\right)^2 \left[\sqrt{1+\left({r\over r_0}\right)^2
} -1\right].
\end{equation}
These profiles have the property that the specific angular momentum $j=\Omega r^2$ 
is a linear function of the enclosed mass.
We choose a central surface density $\Sigma_0=0.12$~g~cm$^{-2}$ 
(corresponding to a central number density $n_0=10^6$ cm$^{-3}$)
and $r_0=c_{\rm s}^2/(1.5 \, G \Sigma_0)$, so that
the latter is comparable to the Jeans length of an isothermal sheet. 
The central angular velocity is $\Omega_0=1.0$~km~s$^{-1}$~pc$^{-1}$.

%
%

\section{Long-term evolution of the mass accretion rate}
\label{longterm}

\begin{figure}
  \resizebox{\hsize}{!}{\includegraphics{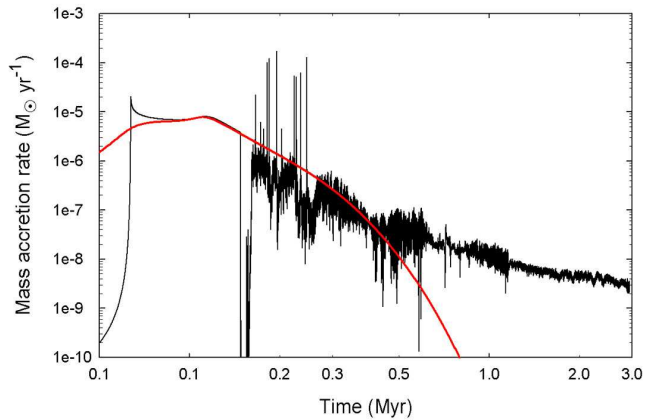}}
      \caption{Temporal evolution of the mass accretion rate through the inner computational boundary
      at $r=5$~AU (solid line) and the mass infall rate on to the disc (red
      line).}
         \label{fig1}
\end{figure}

Our simulations employing the thin-disc approximation can perform efficient calculation of 
the time evolution starting from the prestellar phase and ending
in the T Tauri phase when the central protostar is about 3~Myr old.
Such long-term integration coupled with resolution of a wide range of length
scales is currently prohibitive for three-dimensional simulations.

The solid line in Fig.~\ref{fig1}
shows the temporal evolution of the mass accretion rate 
$\dot{M}(t)=-2\pi r v_r \Sigma(r)$ in to the sink cell,
where we use the inflow velocity $v_r$ of gas at $r=5$~AU.
A sharp increase in the mass accretion rate at $t\approx 0.064$~Myr 
to a peak value $\dot{M}\approx 2.1\times 10^{-5}~M_\odot$~yr$^{-1}$ 
marks the formation of the central 
protostar. The subsequent evolution is characterised by a near-constant accretion at 
$\dot{M}\approx 10^{-5}~M_\odot$~yr$^{-1}$ until $t\approx 0.11$~Myr 
when the rarefaction wave from the outer boundary
approaches the protostar. This phenomenon is explained in detail by \citet{VB3}. 
The mass accretion rate then gradually declines until $t\approx 0.15$~Myr
when the first layer of gas from the infalling envelope hits a centrifugal 
barrier outside the sink cell and the protostellar disc begins to form. During the subsequent
evolution, matter initially lands on the disc rather than falling directly in to
the central sink. The mass accretion rate 
initially drops to a negligible value but rises to $\dot{M}\approx10^{-6}~M_\odot$~yr$^{-1}$
shortly thereafter. The enhanced accretion is 
attributable to newly developed
gravitational instability and spiral structure, induced by the 
continuous infall of matter from the envelope. Dense clumps form within the spiral 
arms and are quickly driven on to the protostar by the action of gravitational torques 
from the spiral arms. These episodes of clump infall produce the bursts of mass accretion 
seen in Fig.~\ref{fig1} between $t\approx 0.17$~Myr and $t\approx 0.25$~Myr.
During the bursts, $\dot{M} \approx 10^{-4}~M_\odot$~yr$^{-1}$, 
which is reminiscent of the FU Ori eruptions.
This ``burst mode'' was discussed in detail by \citet{VB1,VB2}.
We note that the disc mass never exceeds 
$\approx 0.1~M_\odot$, which is reached at 0.45~Myr.
Subsequently, it gradually declines to $0.077~M_\odot$
at 3~Myr. These numbers correspond to disc-to-star mass ratios of 0.15 and
0.11, respectively.

In this paper, we focus on the evolution of the mass accretion rate {\it after} 
the bursts cease to occur. 
Figure~\ref{fig1} indicates that
$\dot{M}$ is in the range $(10^{-7}-10^{-6})~M_\odot$~yr$^{-1}$ in the 
quiescent phase between the bursts and then decreases gradually in the later 
residual accretion phase (after the burst mode has ended), reaching
$\dot{M}\approx 3\times 10^{-9}~M_\odot$~yr$^{-1}$ at $t=3$~Myr.
It is interesting to compare $\dot{M}$ with $\dot{M}_{\rm disc}$, 
the mass accretion rate on to the
disc, calculated at $r=600$ AU. This is safely 
outside the centrifugal disc that is usually localised within the inner 100~AU.
The red line shows the evolution of $\dot{M}_{\rm disc}$. It is 
greater than $\dot{M}$ during the quiescent phase of the burst mode and becomes much less than
$\dot{M}$ after $t\approx 0.5$~Myr. {\it The bursts tend to occur during the time that}
$\dot{M}_{\rm disc}>\dot{M}$. In principle, this inequality can be estimated
observationally and used to determine if a disc is in the quiescent phase 
of the burst mode, i.e. in between mass accretion bursts.

\section{Radial structure of protostellar discs}
\label{radial}

\begin{figure*}
 \centering
  \includegraphics[width=15 cm]{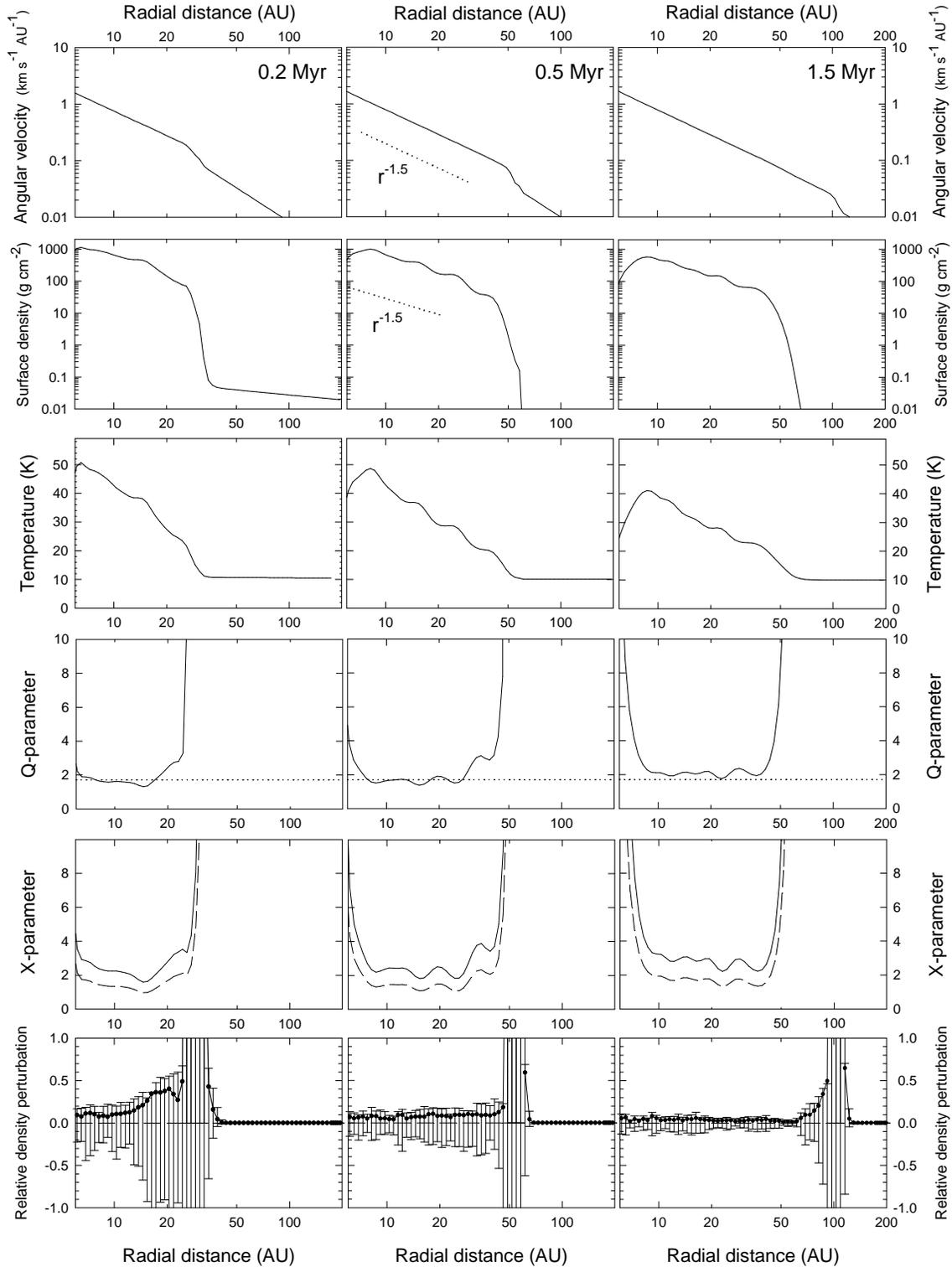}
      \caption{
Top to bottom: azimuthally-averaged values $\overline{\Omega}$, $\overline{\Sigma}$, $\overline{T}$,
$\overline Q$, $\overline X$, and $\overline{\triangle \Sigma}$
at $t=0.2$~Myr (left column), $t=0.5$~Myr (middle column), and $t=1.5$~Myr (right column).
      }
         \label{fig2}
\end{figure*}

%

Insight into the phenomenon that drives accretion in the quiescent
phase between the bursts and in the later residual accretion phase can be obtained from
the radial structure of protostellar discs. Figure~\ref{fig2}
shows the radial profiles of the azimuthally-averaged values of 
$\Omega$, $\Sigma$, temperature $T$, Toomre parameter
$Q$, the $X$-parameter (defined later), and the relative
surface density perturbation $\triangle \Sigma$ 
at $t=0.2$~Myr (left column), $t=0.5$~Myr (middle column), and $t=1.5$~Myr (right column).
The Toomre parameter is calculated as $Q=\tilde{c}_{\rm s} \Omega 
/\pi G \Sigma$, where $\tilde{c}_{\rm s}=(\partial {\cal P}/ \partial
\Sigma)^{1/2}$ is the effective sound speed. 
In each computational zone ($r_i,\phi_j$) we calculate
\begin{equation}
\triangle \Sigma(r_i,\phi_j) = {\Sigma(r_i,\phi_j) - {1\over N} \sum \limits_{j=1}^N \Sigma(r_i,\phi_j)
\over \Sigma(r_i,\phi_j)},
\end{equation}
where $N$ is the number of grid zones in the azimuthal direction. 

Several important conclusions can be drawn by analysing Fig.~\ref{fig2}, which
we review from top to bottom.
The first row shows that the protostellar disc is characterised by 
near-Keplerian rotation during
its evolution, since the stellar mass always dominates the disc mass 
\citep[see fig. 6 of][]{VB2}. (This justifies our use of $\Omega$ in the
definition of $Q$, since the epicyclic frequency $\kappa = \Omega$ for Keplerian
rotation.) Note the visible break in the $\Omega$ profile which marks the edge of the
centrifugal disc. The surface density profile (second row) is also proportional to $r^{-3/2}$ 
in much of the disc, with the exceptions of near the
inner and outer boundaries. 
This profile has a similar form as the minimum mass solar nebula,
$\Sigma = 1000\,r_{\rm AU}^{-3/2}$ g cm$^{-2}$ \citep{Weiden}.
The mild decrease in surface density
near the disc's inner boundary is caused by the
inflow condition on the inner computational boundary. At the same time, the
surface density at several tens of AU shows a sharp drop by many orders
of magnitude, implying that the disc has a sharp physical boundary.
{\it This is an important new result of our self-consistent calculation of disc formation.}
We emphasise that the outer computational boundary is at 8000~AU
and has virtually no influence on the protostellar disc dynamics or the formation
of its sharp edge. The disc
grows in radius from approximately $30$~AU at $0.2$~Myr to about $70$~AU 
at 3~Myr.

The third row shows that the disc, in contrast to the envelope, is decidedly
nonisothermal. As with $\Omega$ and $\Sigma$, there is a sharp drop in temperature at 
the disc edge. The temperature generally increases inward as the surface density
increases, in accordance with the barotropic equation of state.
Nevertheless, the variation in temperature within the disc is much less than
the variations in $\Omega$ and $\Sigma$, which follow power-law profiles.
We note that the temperature distribution in our model disc during the late
accretion phase is an approximate value
since we have not included the complex roles of stellar irradiation and radiative
transfer. The model of \citet{Chiang}, which accounts for heating from a 
central star onto a passive minimum mass solar nebula model, yields temperatures 
in the range 50-60 K for a disc at 10 AU, in comparison to our model temperature 
of 40 K at this radius at late times. More detailed models by \citet{Garaud},
who add viscous dissipation within the context of steady-state accretion 
in an $\alpha$-prescription (the resulting surface density is much shallower
than $r^{-3/2}$), yield even greater midplane temperatures in 
many cases, but generally agree with the \citet{Chiang} model for
an accretion rate 10$^{-8}$ \msun~yr$^{-1}$. An exact comparison of our
calculated temperatures with either of these models is difficult due to their
sensitive dependence on disc flaring, a quantity that we do not follow 
in this study, and due to the different surface density distribution of
our model discs.

The radial distribution of the azimuthally-averaged Toomre parameter $\overline{Q}$ 
is shown in the fourth row of Fig.~\ref{fig2}. 
A substantial portion of the disc
at $t=0.2$~Myr and $t=0.5$~Myr is characterised by $\overline Q$ below a 
fiducial critical value
of $\sqrt 3$ \citep[estimated from the linear analysis of][and shown
by the dotted line]{Polyachenko}. 
However, at $t=1.5$~Myr $\overline Q$ becomes greater
than $\sqrt 3$ and the relative amplitude of density perturbations in the disc is
expected to diminish. In all cases, $\overline{Q}$ maintains near-uniformity in 
space due to a self-regulation mechanism in the disc that prevents it from falling
substantially below the critical value.  
If $Q$ falls below a critical value 
$\Qcr$, the gas disc becomes vigorously unstable, develops
a spiral structure, and may even fragment to form dense clumps within the arms.
Accretion on to the central sink as well as heating induced by density enhancements
work to restore $Q$ back toward the critical value.
Ultimately, the near $r^{-3/2}$ surface density profile within the disc that is 
illustrated in the second row is a result of this self-regulation of the Toomre
parameter coupled with the near-Keplerian angular velocity and relatively weak
temperature variation.
We note that our obtained distribution of the azimuthally
averaged Toomre parameter  is
qualitatively similar to that obtained by \citet{Lodato04}
using SPH simulations of isolated discs with a simple parametrisation for the cooling function.


Although $Q$ cannot drop to arbitrarily {\it low} values, what prevents the disc from 
quickly achieving a {\it high} $Q$ state in which
density fluctuations are quickly washed out due to 
pressure gradients or shear? 
We believe that the answer is swing amplification.
Numerical simulations indicate that the transition between stable
and unstable phases is usually a smooth process rather than an instantaneous switch.
There can exist a situation when the disc sustains low-amplitude 
density perturbations for a substantially long time even though the Toomre
parameter averaged over the whole disc is above the critical value and
the disc is stable {\it globally}. This phenomenon can take place if gravitational instability is 
powered by swing amplification. Amplification occurs when any leading spiral disturbance unwinds 
into a trailing one due to differential rotation \citep[][and others]{GLB,Toomre}.
The importance of swing amplification is in its essentially local nature -- it provides
transitory gravitational amplification to a local patch of gas where a leading spiral disturbance 
unwinds into a trailing one and the stabilising influence of shear is temporarily cancelled 
\citep[][p. 378]{BT}.
It is important to realise that swing amplification can work when the disc is stable globally 
but unstable {\it locally}.   
In this case, spiral density perturbations would be amplified locally
(if local conditions favour gravitational instability) but the gain would be too small
for gravitational instability to grow globally throughout the disc. 
In order for this process to work for a long time, a feedback mechanism must be present that
constantly feeds a gas disc with leading spiral disturbances \citep[][p. 379]{BT}. In this case,
a reflection of trailing spiral disturbances from the sharp outer disc edge 
can provide the necessary feedback mechanism.

It is convenient to introduce the $X$-parameter when considering the effect
of swing amplification.  It is defined as 
$X \equiv \lambda/\lambda_{\rm cr}$, where 
$\lambda \equiv 2\pi r/m $ is the circumferential wavelength of an $m$-armed spiral disturbance,
$\lambda_{\rm cr} \equiv 4\pi^2 G \Sigma/\kappa^2$ is the longest unstable wavelength 
in a cold disc, and $\kappa$ is the epicyclic frequency.
According to \citet{GLB}, the gain from the 
swing amplifier in a gaseous disc is greatest when $0.5\la X \la 2.5$.  
Generally speaking, swing amplification is most efficient when 
$\lambda \approx \lambda_{\rm cr}$. For $\lambda \gg \lambda_{\rm cr}$ and $\lambda \ll \lambda_{\rm cr}$, 
the swing amplifier is strongly moderated by local shear and gas pressure, respectively.
Figure~\ref{fig2} (fifth row)
shows the radial distribution of the azimuthally-averaged 
$X$-parameter in the disc for the $m=6$
spiral mode (solid line) and $m=10$ spiral mode (dashed line). 
It is evident that the favourable conditions for swing amplification are
present in the inner few tens of AU and the maximum gain of the swing amplifier 
is expected for higher order modes $m \approx 10$. Consequently, the disc is expected to develop 
a flocculent multi-armed spiral structure, which is indeed seen in fig.~4 of \citet{VB2}.

The presence of swing-amplified density perturbations is illustrated
in Fig.~\ref{fig2} (bottom row). The filled circles show the azimuthally-averaged 
relative density perturbations $\overline{|\triangle \Sigma|}$
at different disc radii. Note that we azimuthally average the absolute
values of $\triangle \Sigma$ in order to avoid a partial cancellation of
positive and negative relative density perturbations along the azimuth.
The error bars indicate the maximum positive and negative relative
density perturbations at a specific radius. 
A substantial portion of the disc at $t=0.2$~Myr is characterised by $0.1<\overline{\triangle
\Sigma}<0.5$. A steady increase of $\overline{\triangle
\Sigma}$ with radius is evident, suggesting that density perturbations are powered by the swing-amplified
spiral disturbances reflected off the disc outer edge. 
The magnitude of relative density perturbations decreases with time,
which agrees with a gradual stabilisation of the disc inferred from the
temporal evolution of $\overline{Q}$.
For instance,  $\overline{\triangle \Sigma}\approx 0.1$ throughout most of the disc at 
$t=0.5$~Myr, though local
relative density perturbations can fall within the range $[+0.2,-0.3]$.
At $t=1.5$~Myr,  $\overline{\triangle\Sigma}\approx 0.01$ everywhere except just
at the outer disc edge. We note that the actual densities are very low
there and large relative density perturbations have little  influence on the disc dynamics. 


%
To summarise, the nonaxisymmetry diminishes with time and is less pronounced
in $10^6$ yr-old discs than in $10^5$ yr-old discs. 
Interestingly, nonaxisymmetric structure
is observed in the several Myr
old discs around HD 100546 and AB Aurigae \citep{Grady,Fukagawa}.

\begin{figure}
  \resizebox{\hsize}{!}{\includegraphics{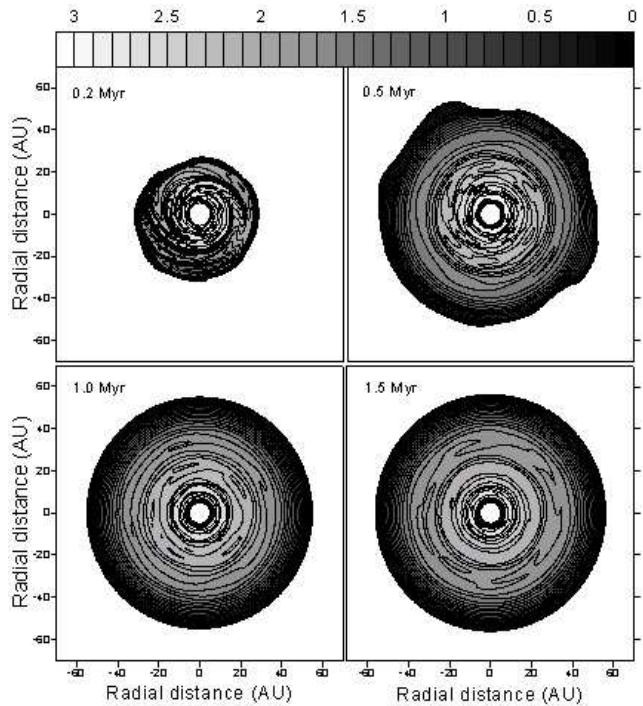}}
      \caption{Surface density distribution in the protostellar disc at
      four evolutionary times indicated in each frame. The scale bar is
      in g~cm$^{-2}$. The gas with surface density below 1.0~g~cm$^{-2}$
      is shown with white space. The central white circle represents the
      protostar plus some circumstellar matter that is 
      unresolved in our numerical simulations. }
         \label{fig3}
\end{figure}

\begin{figure}
  \resizebox{\hsize}{!}{\includegraphics{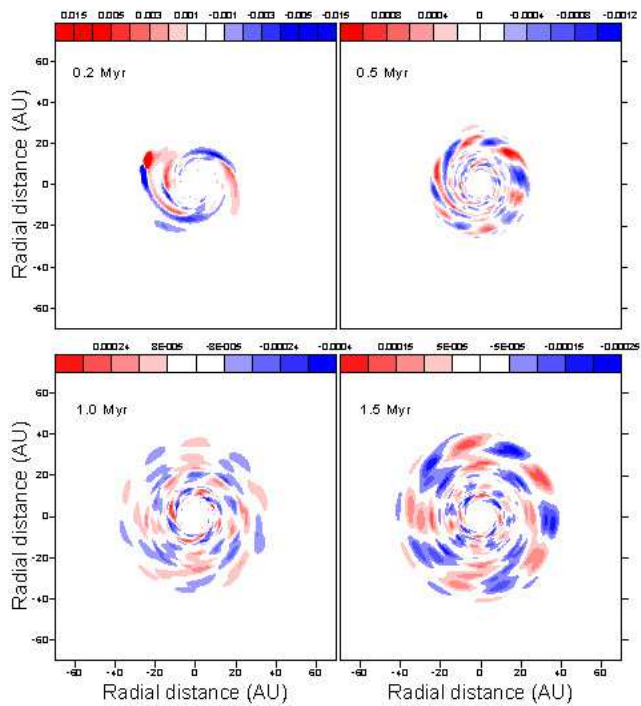}}
      \caption{Spatial distribution of gravitational torques at four evolutionary
      times corresponding to the surface density distributions in Fig.~\ref{fig3}.
      The positive and negative gravitational torques are shown with the
      shades of red and blue, respectively. The scale bars are in 
      units of $8.66\times 10^{40}$~g~cm$^2$~s$^{-2}$. }
         \label{fig4}
\end{figure}

\section{A closer look at gravitational torques}
\label{torque}

\subsection{Self-regulated gravitational accretion}
\label{torque1}

Even the slightest deviation from axial symmetry in a protostellar disc
produces gravitational torques. The strength of the gravitational torques 
contributes to the rate of angular momentum redistribution \citep{Larson}, 
and by implication, the mass accretion rate. In this section, we 
consider the effect of gravitational torques on the radial redistribution of mass
and angular momentum within protostellar discs.

Figure~\ref{fig3} illustrates the gas surface density of the protostellar
disc at four evolutionary times \citep[see also fig.~4 in][]{VB2}.
It is evident that multiple spiral arms are
present at $t=0.2$~Myr but that they disappear by about $t=0.5$~Myr. During the
subsequent evolution, only low-amplitude nonaxisymmetric density 
inhomogeneities are present in the disc.
It is well known that negative torques drive matter inward and 
angular momentum outward, and that positive torques act in the opposite 
manner \citep[e.g.][and others]{Laughlin,Tomley,VT}.
A spatial distribution of gravitational torques in the protostellar disc
is shown in Fig.~\ref{fig4}, in which the positive and negative gravitational
torques $\tau(r,\phi)$ are plotted at four evolutionary times in shades of red 
and blue, respectively \citep[see also fig.~5 in][]{VB2}. The regions with near-zero 
$\tau(r,\phi)$ are shown 
with white space. 
Trailing spiral arms can be clearly seen in the distribution of
gravitational torques at $t=0.2$~Myr. In particular, the inner part of a spiral arm is usually 
characterised by negative $\tau(r,\phi)$, whereas the end of a spiral is characterised by positive
$\tau(r,\phi)$. 

\begin{figure}
  \resizebox{\hsize}{!}{\includegraphics{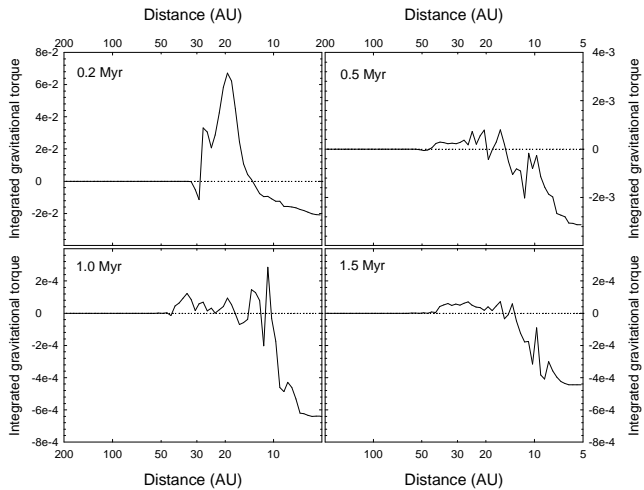}}
      \caption{Cumulative gravitational torques ${\cal T}(r)$ as a 
      function of radial distance $r$ (see text for explanation)
      at four evolutionary times. 
      Note that distance decreases to the right.}
         \label{fig5}
\end{figure}

\begin{figure}
  \resizebox{\hsize}{!}{\includegraphics{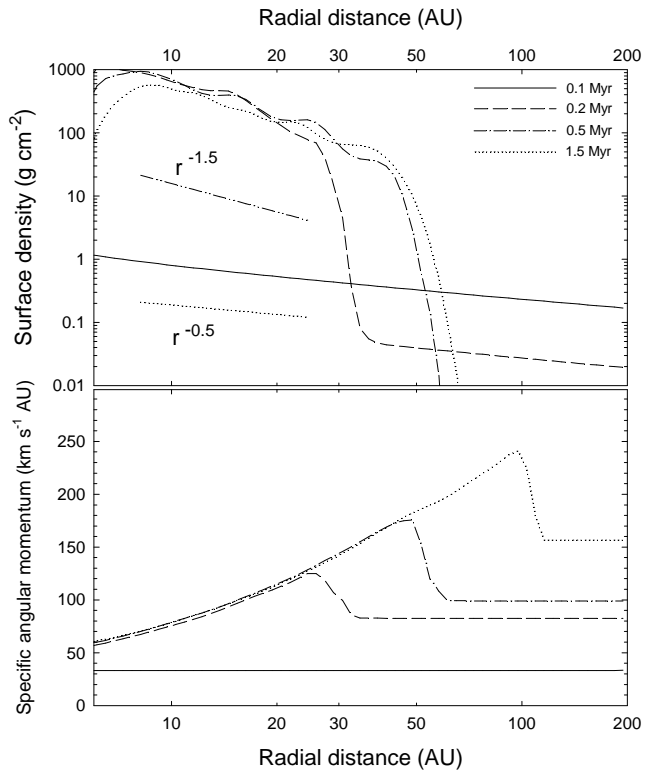}}
      \caption{Radial profiles of gas surface density (top) and specific
      angular momentum (bottom) at four evolutionary times. The disc forms
      at $t\approx 0.15$~Myr. The inward transport of mass and the outward
      transport of specific angular momentum after the disc formation results
      in a steepening of the radial profiles of the gas surface density
      and specific angular momentum. }
         \label{fig6}
\end{figure}

The spatial distribution of gravitational torques in the later evolution ($t
\ga 0.5$~Myr) is particularly interesting. It shows an alternating behaviour
in the azimuthal direction -- local gas patches characterised by positive $\tau(r,\phi)$ 
are followed by patches with negative $\tau(r,\phi)$ and vice versa.  
We note that this pattern fluctuates with time.
Even though the net global effect of negative 
and positive torques appears (by eye) to cancel out, it is clear that the
net inward mass accretion exhibited by our model requires a residual
negative gravitational torque (due to the density inhomogeneities), 
at least within the inner disc.  

To further investigate the net effect of 
local positive and negative gravitational torques when averaged over a
significant portion of the disc,
we calculate the cumulative gravitational torque
${\cal T}(r)$ in a radial bin between 200~AU and $r$. For instance, ${\cal T}(r=30~{\rm
AU})$ would represent the sum of all local gravitational torques $\tau(r,\phi)$ starting from 200~AU and ending at $r=30$~AU.
The gravitational torques are expected
to be small at large radial distances due to both the low gas surface density 
and effective axial symmetry in the envelope. Figure~\ref{fig5} shows the resulting integrated 
gravitational torque as a function of radial distance at four evolutionary times.  
Clearly, it is virtually zero at large 
radial distances. Moving radially inward, the integrated gravitational
torque first becomes positive and then becomes negative. 
This implies that the outer parts of the protostellar disc are 
dominated by {\it positive} gravitational torques, whereas the inner regions
of the disc are dominated by {\it negative} gravitational torques. 
This important feature contributes to the overall 
mass and angular momentum redistribution within the disc.

The global effect of gravitational torques is illustrated in Fig.~\ref{fig6},
where we show the radial profiles of the gas surface density $\Sigma$ (top) and specific 
angular momentum $j=\Omega r^2$ (bottom) at four evolutionary times. 
Before disc formation at $t\approx 0.15$~Myr, the infalling envelope has 
a shallow radial surface density profile, essentially 
$\propto r^{-0.5}$, and the specific angular momentum is nearly 
uniform in the inner 200~AU.
Profiles of the form $\Sigma \propto r^{-0.5}$ and $\Omega \propto r^{-2}$
($j = $ const.) are the expected self-similar profiles for 
the freely-falling regions inside an expansion wave but outside the
centrifugal disc \citep{Saigo}. 
After disc formation, the radial density within the disc develops
a much steeper profile, 
$\Sigma\propto r^{-1.5}$, indicating an inward transport of mass. 
On the other hand, the specific angular momentum in the disc 
is obviously transported outward, 
which is indicated by a developed positive slope in the radial
distribution of $j$. Outside the disc, the specific angular momentum 
remains nearly spatially uniform though growing with time. 

\subsection{Diffusive nature of the accretion process}
\label{torque2}

A diffusive nature of the radial mass transport in
the late evolution of the disc can be visualised by considering the temporal evolution of a radially narrow 
gas annulus located initially at some distance from the inner disc boundary. 
Any mechanism
of radial mass and angular momentum transport will change the radial 
position and shape of the annulus during the subsequent evolution. 

\begin{figure}
  \resizebox{\hsize}{!}{\includegraphics{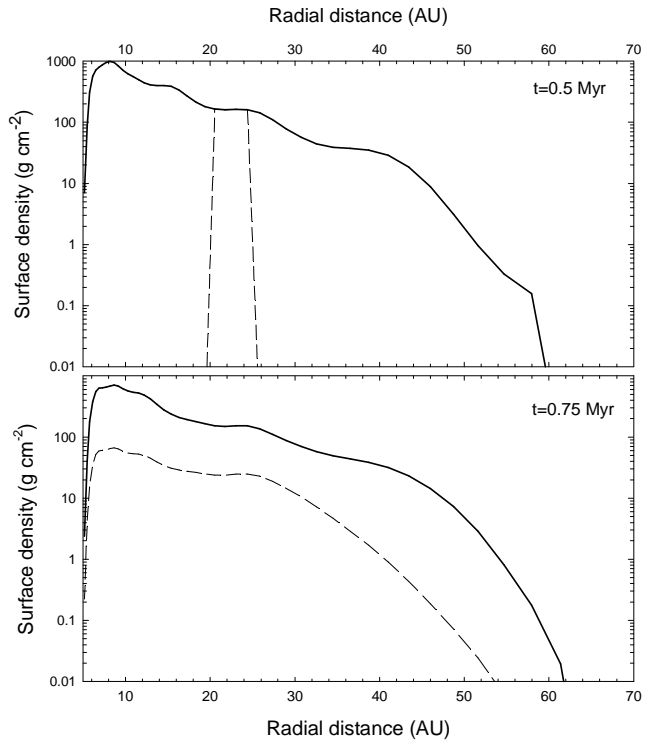}}
      \caption{Gravitationally-driven diffusion of a radial annulus.
      Top panel: 
      the disc surface density distribution (solid line) and 
      the surface density in an annular region (dashed line)
      at $t=0.5$~Myr. Bottom panel:
      the disc surface density (solid line) and the surface density of mass previously
      in the annulus, at $t=0.75$~Myr. 
      The mass originally in the annulus has spread out rather than 
      been translated homologously, and has a radially
      declining surface density profile. This indicates a 
      predominantly inward mass transport.
      }
         \label{fig7}
\end{figure}

To perform this test, we
follow the evolution of a radial gas annulus initially located between 20~AU and 25~AU.
We solve a separate continuity equation for the annulus starting at $t=0.5$~Myr,
when the protostellar disc has reached a radial size of approximately 50~AU. 
The initial gas surface density of the annulus is shown in the upper panel
of Fig.~\ref{fig7} with the dashed line. It equals the corresponding gas
surface density in the disc and is set to a negligibly small value
elsewhere. The resulting gas surface
density distribution of the annulus at $t=0.75$~Myr is shown in the bottom
panel of Fig.~\ref{fig7} with the
dashed lines. For comparison, the solid line shows the gas surface density in the
disc at the same evolutionary time. Clearly, the annulus
spreads out rather than moves homologously in the radial direction. 
This behaviour is in agreement with a diffusive nature
of the radial mass transport in a protostellar disc.
Because the net gravitational torque is negative, the annulus
spreads out predominantly  inward, resulting in a radially declining surface density
profile at $t=0.75$~Myr. To quantify this effect, we find the percentile of 
the initial mass of the annulus ($1.04\times 10^{-2}~M_\odot$) that was transported 
inwards and outwards of its initial location between 20~AU and 25~AU. 
At $t=0.75$~Myr, the mass inside 20~AU is $5.2\times10^{-3}~M_\odot$ or 50 per cent 
of the initial annulus mass and the mass outside 25~AU is $3.1\times 10^{-3}~M_\odot$ or
30 per cent of the initial annulus mass. These percentiles show an even 
greater contrast during the subsequent evolution at $t>0.75$~Myr.
Another intriguing result is that the shape of the gas surface 
density profile in the
spread-out annulus (dashed line) closely resembles that of the disc (solid line) 
after $t=0.75$~Myr. 

\section{Summary and Discussion}

Our model of the self-consistent formation of a protostellar 
disc from the collapse of its parent cloud core and the subsequent
long-term evolution of the disc reveals that it settles into
a self-regulated state. In this state, the gravitational torques
are responsible for inward mass transport 
at a rate that is in agreement with typical accretion rates 
inferred in T Tauri star discs \citep{Hartmann98b}. 
Persistent nonaxisymmetric density perturbations are the key
ingredient that leads to the gravitational torques. 
The amplitude of these nonaxisymmetric density perturbations
decreases with time but the decline is moderated by the 
action of swing amplification, aided by the presence of a
self-consistently-formed sharp edge at the interface between the
protostellar disc and core envelope. The disc self-regulation
by gravitational torques maintains a near-uniform value of the
azimuthally-averaged Toomre $Q$ parameter. The mass accretion
also maintains the disc-to-star mass ratio to be $\approx 0.1$,
and the rotation rate within the disc is approximately Keplerian.
Since the temperature variation in the disc is much smaller in
magnitude than the power-law variation in $\Omega$, the 
self-regulated state with near-uniform $Q$ implies that
the surface density $\Sigma \propto r^{-3/2}$.

The near uniformity of $Q$ may be understood using an analogy
to convection in stellar envelopes. A highly superadiabatic
temperature gradient cannot be maintained in a stellar envelope,
since convection will set in and bring the
temperature gradient back to a near-adiabatic value. Similarly,
gravitational torques work to maintain $Q$ at essentially the 
stability limit. However, it must be emphasised that {\it
our model discs are not turbulent}. There are  
density perturbations in the disc, but they are not fed in 
globally from the largest scales of the disc and are indeed declining in
amplitude over time. The swing amplification mechanism does
sustain the structure for a substantially long time, but is 
driven by transient local gravitational instabilities. 
We do not believe that the labels ``turbulent'' or even 
``gravitoturbulent'' are justified: rather we refer to this mode
as ``self-regulated gravitational accretion''.

The $\Sigma \propto r^{-3/2}$ profile is in agreement
with estimates of the minimum mass solar nebula \citep{Weiden}, made
by adding the solar abundance of light elements to each planet and
spreading the masses through zones surrounding the planetary orbits.
The disc-to-star mass ratio in our model remains greater than 
that inferred for the solar nebula or for typical external star-disc
systems. However, \citet{Hartmann}
have recently pointed out that disc masses based on dust emission
may be systematically underestimated. In any case, other transport
mechanisms, primarily associated with magnetic fields, may indeed
operate in real discs, and may also be more important in the inner
5 AU that we do not model. 
Since the level of magnetic coupling in the disc also remains 
uncertain, our model provides a very clear 
physical baseline for the global behaviour of discs before 
accounting for other more poorly-understood effects.

 
\label{summary}

\section*{Acknowledgments} 
We thank the referee and Wolfgang Dapp for helpful comments on 
the manuscript, and 
Martin Houde and the SHARCNET consortium for access to 
computational facilities.
This work was supported by the Natural Sciences and Engineering
Research Council of Canada, RFBR grant 06-02-16819-a, South Federal
University grant 05/6-30, and Federal Agency of Education 
(project code RNP 2.1.1.3483).  EIV gratefully acknowledges support 
from a CITA National Fellowship

\appendix 
\section{The Relative Role of Artificial Viscosity}

\subsection{Comparison with gravitational torques}

\begin{figure}
  \resizebox{\hsize}{!}{\includegraphics{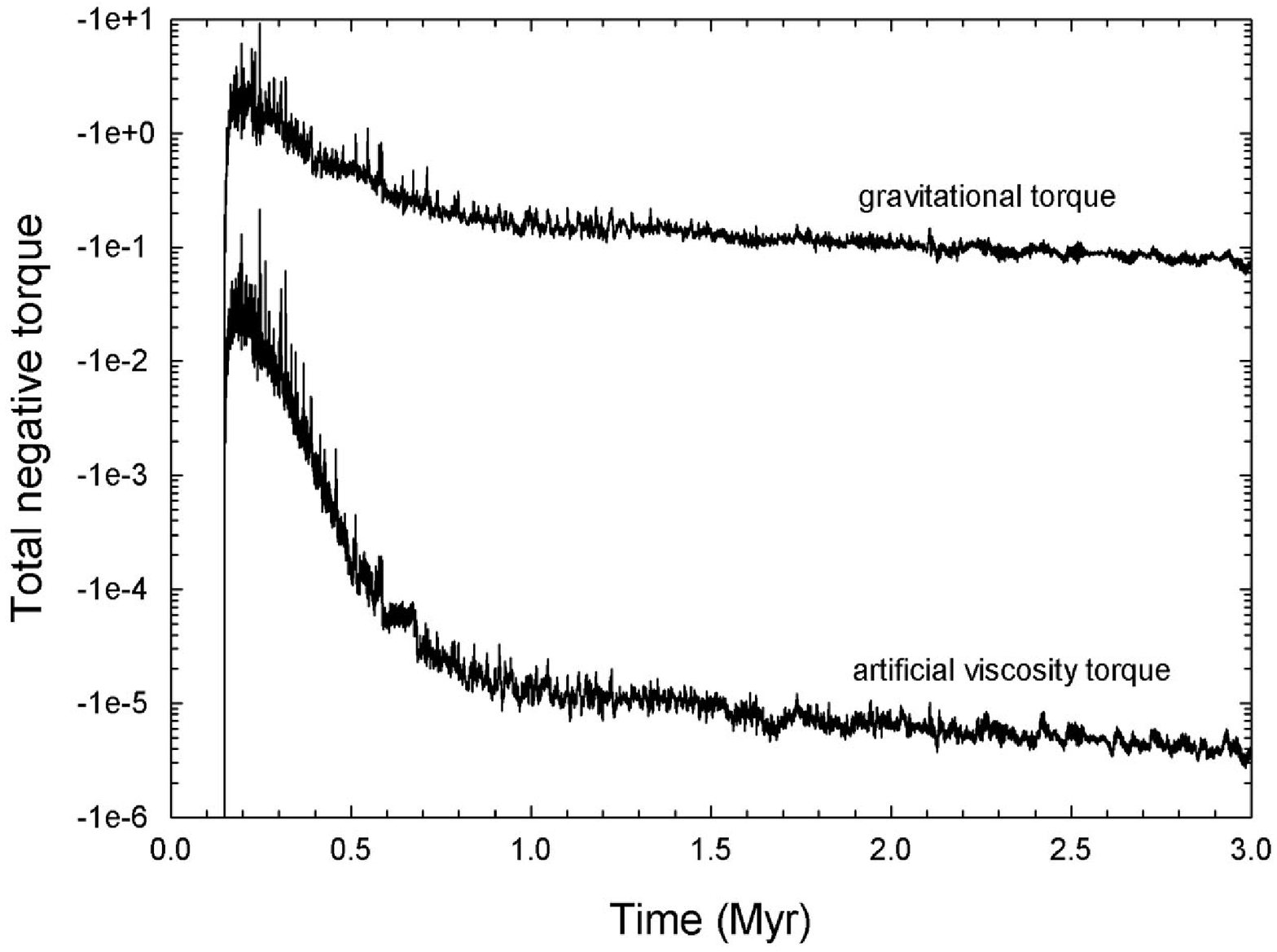}}
      \caption{Comparison of the total negative torques due to gravity (upper
      curve) and artificial
      viscosity (lower curve). Note that the total negative gravitational
      torque is much greater than the total negative artificial viscosity
      torque during the evolution.}
         \label{fig8}
\end{figure}

In principle, the torques due to artificial viscosity (present in our Eulerian numerical 
hydrodynamics code) could also contribute to the radial transport of mass and angular momentum.
To quantify this effect, we calculate 
the net artificial viscosity torque in the disc ${\cal T}_{\rm av}$, defined as the sum of 
local artificial viscosity torques 
$\tau_{\rm av}(r,\phi)=-S(r,\phi)\, \partial P_{\rm av}/\partial \phi$ in the inner 600~AU.
Here, $S(r,\phi)$ is the surface area occupied by a cell with polar coordinates $(r,\phi)$
and $P_{\rm av}$ is the artificial viscosity pressure defined according to the usual 
von Neumann \& Richtmyer prescription \citep[see e.g.][]{SN}.
The net gravitational torque ${\cal T}_{\rm gr}$ is
found by summing all local gravitational torques $\tau(r,\phi)$ in the inner 600~AU.
We find that ${\cal T}_{\rm av}$ is at least ten orders of magnitude smaller than ${\cal T}_{\rm
gr}$ and fluctuates near zero during the evolution, meaning that the positive and 
negative artificial viscosity torques essentially cancel each other globally. 
Furthermore, we must make sure that 
even the negative part of the artificial viscosity torque cannot compete 
with the corresponding negative gravitational torque and lead to 
spurious radial mass transport.
Hence, we also calculate the total {\it negative} torques due to gravity and artificial viscosity
by summing only negative values of $\tau(r,\phi)$ and $\tau_{\rm av}(r,\phi)$ in the inner 600~AU, 
respectively. Figure~\ref{fig8} shows these total negative torques at different evolutionary times.
The negative torque due to gravity is clearly many orders of magnitude 
greater than the negative torque due to artificial viscosity, especially in the
late disc evolution when the spiral arms have disappeared. These tests enable us to conclude
that artificial viscosity 
has little influence on the radial transport of mass in our numerical simulations.

\subsection{Accretion without gravity}

\begin{figure}
  \resizebox{\hsize}{!}{\includegraphics{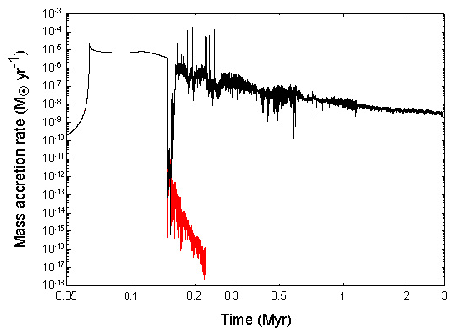}}
      \caption{Temporal evolution of the mass accretion rates $\dot{M}$
      with (black line) and without (red line) self-gravity
      of the disc.  Before  the disc formation at $t\approx 0.17$~Myr
      both accretion rates are identical. After the disc formation, the
      mass accretion rate in the non-self-gravitating disc declines quickly
      to a negligibly small value.} 
         \label{fig9}
\end{figure}

A decisive test to confirm the principal role of gravitational torques is to 
see if a non-self-gravitating disc can drive accretion rates comparable to those
in the self-gravitating disc. To do this, we turn off self-gravity at $t=0.15$~Myr,
i.e. just before disc formation but well after the central protostar has formed.
Figure~\ref{fig9} shows the previously calculated $\mdot$ with self-gravity (black line) as well
as the newly computed $\mdot$ in the absence of self-gravity (red line).
The two rates are identical before the disc forms at $t\approx 0.17$~Myr.
Afterwards, $\mdot$ in the non-self-gravitating
disc declines quickly to a remarkably low value 
$\dot{M}\approx 10^{-17}~M_\odot$~yr$^{-1}$ during 0.1~Myr, 
confirming the principal role of self-gravity and gravitational 
torques in driving the mass accretion in our model.
We also note that it is not possible to follow the simulations 
further in time, since the gas density near the inner boundary becomes negative. 
This is because the accretion is then driven by small numerical imperfections of the inner 
inflow computational boundary. As a result, the computational cells that are immediately
adjacent to the inner boundary can become depleted of gas due to the absence of 
gravitationally driven inward transport of gas from the outer disc. 


\begin{thebibliography}{}



\bibitem[\protect\citeauthoryear{Andrews \& Williams}{2005}]{Andrews}
Andrews, S. M., Williams, J. P., 2005, ApJ, 631, 1134 

\bibitem[\protect\citeauthoryear{Balbus \& Hawley}{1998}]{Balbus}
Balbus, S. A., Hawley, J. F., 1998, Rev. Mod. Phys., 70, 1

\bibitem[\protect\citeauthoryear{Basu}{1997}]{Basu}
Basu, S., 1997, ApJ, 485, 240  

\bibitem[Binney \& Tremaine(1987)]{BT}
Binney, J., \& Tremaine, S., 1987, Galactic Dynamics. 
Princeton Univ. Press, Princeton

\bibitem[\protect\citeauthoryear{Chiang \& Goldreich}{1997}]{Chiang}
Chiang, E. I., Goldreich, P., 1997, ApJ, 490, 368



\bibitem[\protect\citeauthoryear{Fukagawa et al.}{2004}]{Fukagawa}
Fukagawa, M., Hayashi, M., Tamura, M., Itoh, Y. et al., 2004, ApJ, 605, L53

\bibitem[\protect\citeauthoryear{Gammie}{1996}]{Gammie96}
Gammie, C. F., 1996, ApJ, 457, 355


\bibitem[\protect\citeauthoryear{Garaud \& Lin}{2007}]{Garaud}
Garaud, P., Lin, D. N. C., 2007, ApJ, 654, 606

\bibitem[\protect\citeauthoryear{Glassgold et al.}{1997}]{Glassgold}
Glassgold, A. E., Najita, J., Igea, J., 1997, ApJ, 480, 344

\bibitem[\protect\citeauthoryear{Goldreich \& Lynden-Bell}{1965}]{GLB}
Goldreich, P., Lynden-Bell, D., 1965, MNRAS, 130, 125

\bibitem[\protect\citeauthoryear{Goodman et al.}{1993}]{Goodman}
Goodman, A. A., Benson, P. J., Fuller, G. A., Myers, P. C.,
1993, ApJ, 406, 528

\bibitem[\protect\citeauthoryear{Grady et al.}{2001}]{Grady}
Grady, C. A., Polomski, E. F., Henning, Th., Stecklum, B., et al., 2001,
AJ, 122, 3396 

\bibitem[\protect\citeauthoryear{Hartmann}{1998}]{Hartmann98b}
Hartmann, L., 1998, Accretion Processes in Star Formation.
Cambridge Univ. Press, Cambridge

\bibitem[\protect\citeauthoryear{Hartmann et al.}{2006}]{Hartmann}
Hartmann, L., D'Alessio, P., Calvet, N., Muzerolle, J., 2006, ApJ, 648, 484



\bibitem[\protect\citeauthoryear{Larson}{1984}]{Larson}
Larson, R. B., 1984, MNRAS, 206, 197

\bibitem[\protect\citeauthoryear{Laughlin \& Bodenheimer}{1994}]{Laughlin}
Laughlin, G., Bodenheimer, P., 1994, ApJ, 436, 335
%

\bibitem[\protect\citeauthoryear{Lodato \& Rice}{2004}]{Lodato04}
Lodato, G., Rice, W. K. M., 2004, MNRAS, 351, 630

\bibitem[\protect\citeauthoryear{Lodato \& Rice}{2005}]{Lodato05}
\ul., 2005, MNRAS, 358, 1489

\bibitem[\protect\citeauthoryear{Lodato et al.}{2007}]{Lodato07}
Lodato G., Meru F., Clarke C. J., Rice W. K. M., 2007, MNRAS, 374, 590


%



\bibitem[\protect\citeauthoryear{Pickett et al.}{2000}]{Pickett}
Pickett, B. K., Cassen, P., Durisen, R. H., Link, R., 2000, ApJ, 529, 1034 

\bibitem[\protect\citeauthoryear{Polyachenko et al.}{1997}]{Polyachenko} 
     Polyachenko V. L., Polyachenko E. V., Strel'nikov A. V., 1997, 
     Astron.\ Zhurnal, 23, 598 (translated Astron.\ Lett.\ 23, 525) 
     


\bibitem[\protect\citeauthoryear{Saigo \& Hanawa}{1998}]{Saigo}
Saigo, K., Hanawa, T., 1998, ApJ, 493, 342

%
\bibitem[\protect\citeauthoryear{Scholz et al.}{2006}]{Scholz1}
Scholz, A., Jayawardhana, R., Wood, K., 2006, ApJ, 645, 1498


\bibitem[\protect\citeauthoryear{Stone \& Norman}{1992}]{SN}
Stone, J. M., Norman, M. L., 1992, ApJS, 80, 753

\bibitem[\protect\citeauthoryear{Tomley et al.}{1991}]{Tomley}
Tomley, L., Cassen, P., Steiman-Cameron, T., 1991, ApJ, 382, 530

\bibitem[\protect\citeauthoryear{Toomre}{1981}]{Toomre}
     Toomre, A., 1981, in Fall S. M., Lynden-Bell D., eds, The Structure
     and Evolution of Normal Galaxies. Cambridge Univ. Press,
     Cambridge

\bibitem[\protect\citeauthoryear{Turner et al.}{2007}]{Turner}
Turner, N. J., Sano, T., Dziourkevitch, N., 2007, ApJ, 659, 729

\bibitem[\protect\citeauthoryear{Vorobyov \& Basu}{2005a}]{VB3}
Vorobyov, E. I., Basu, S., 2005a, MNRAS, 360, 675

\bibitem[\protect\citeauthoryear{Vorobyov \& Basu}{2005b}]{VB1}
\ul., 2005b, ApJ, 633, L137

\bibitem[\protect\citeauthoryear{Vorobyov \& Basu}{2006}]{VB2}
\ul., 2006, ApJ, 650, 956

\bibitem[\protect\citeauthoryear{Vorobyov \& Theis}{2006}]{VT}
Vorobyov, E. I., Theis, Ch., 2006, MNRAS, 373, 197

\bibitem[\protect\citeauthoryear{Weidenschilling}{1977}]{Weiden}
Weidenschilling, S. J., 1977, ApSS, 51, 153


\end{thebibliography}
\end{document}